\input harvmac

\Title{\vbox{\baselineskip12pt\hbox{CERN-TH/96-319} \hbox{UGVA-DPT 1996/10-995}
\hbox{hepth/9612103 }}}
{\vbox{\centerline
{ Confining Strings     
 with Topological Term}}}
 \centerline{M. C. Diamantini $^a$, F. Quevedo $^a$ and C. A. Trugenberger $^b$}
 \bigskip
\centerline{$^a$ CERN Theory Division}
\centerline{CH-1211 Geneva 23,
Switzerland}\medskip
\centerline{$^b$ Department of Theoretical Physics}
\centerline{
University of Geneva,}
\centerline{CH-1211 Geneva 4, Switzerland}

\vskip .3in
\noindent
We consider several aspects of `confining strings', recently
proposed to describe the confining phase of gauge field theories.
 We perform the exact duality transformation
that leads to the confining string action and show that it reduces to
the Polyakov action in the semiclassical approximation.
In 4D we introduce a `$\theta$-term' and compute the low-energy
effective action for the confining string in a derivative 
expansion. We find that the coefficient of the extrinsic
curvature (stiffness) is negative, confirming previous proposals.
In the absence of a $\theta$-term, the effective string action 
is only  a cut-off theory  for finite values of the
coupling $e$, whereas for generic values of $\theta$, the action
can be renormalized and to leading order we obtain the
Nambu-Goto action plus a topological `spin' term that 
could stabilize the system.
\Date{\vbox{\baselineskip12pt\hbox{CERN-TH/96-319}
\hbox{December 1996}}}

\vfill
\eject
\newsec{Introduction}
Despite many efforts the formulation of a consistent string theory for the
confining phase of gauge theories remains an open problem. In order to cure
the diseases of the bosonic string in 4D Polyakov \ref\pol{A.M. Polyakov, Nucl.
Phys. B268 (1986) 406.} \ref\amp{For a review see: A.M. Polyakov, ``Gauge Fields
and Strings'', Harwood Academic Publishers, Chur  (1987).}\ and Kleinert
\ref\kle{H. Kleinert, Phys. Lett. B174 (1986) 335; Phys. Rev. Lett. 58 (1987)
1915.}\ proposed to add to the Nambu-Goto action a term proportional to the
{\it extrinsic curvature} of the world-sheet.
However, the so obtained {\it rigid\ string} is not problem-free.
The inverse of the coupling constant of the new term is asymptotically free
\pol\ \kle, so that, in absence of a non-trivial fixed point, the extrinsic
curvature becomes irrelevant in the infrared.
Moreover, the rigid string suffers from the stability problems common to all
higher-derivative actions and can be viewed at most as a long-distance
expansion of a non-local action \ref\bra{E. Braaten and C.K. Zachos, Phys. Rev.
D35 (1987) 1512; J. Polchinski, UTTG 16-92.}.

A related problem regards the sign of the rigidity term. Clearly, this was
originally proposed to enter the (Euclidean) action with a positive coefficient
in order to suppress strongly creased surfaces.
However, most computations of the extrinsic curvature term for Nielsen-Olesen
vortices \ref\hkl{H. Kleinert, Phys. Lett. B211 (1988) 151; Phys. Lett. B246
(1990) 127; K.I. Maeda and N. Turok, Phys. Lett. B202 (1988) 376; S.M. Barr and
D. Hochberg, Phys. Rev. D39 (1989) 2308; P. Orland, Nucl. Phys. B428 (1994)
221.}\ produced the opposite, negative sign.
No rigidity term was found in \ref\gre{R. Gregory, D. Haws and D. Garfinkle,
Phys. Rev. D42 (1990) 343; R. Gregory, Phys. Rev. D43 (1991) 520.}, while a
positive coefficient was obtained in \ref\rgr{R. Gregory, Phys. Lett. B206
(1988) 199.}.

Recently there were new developments in this field. First, Kleinert
\ref\hke{H. Kleinert, hep-th/9601030.}\ proposed that the extrinsic curvature
{\it should} indeed enter the (Euclidean) action with a negative coefficient
and that this actually improves the stability of the theory.
Secondly, Polyakov \ref\mpo{A.M. Polyakov, PUPT-1632, hep-th/9607049.}\
proposed a new  string action, which he called the {\it confining\
string}.

The (Euclidean) action of the confining string is given by
\eqn\csa{e^{-S_{\rm CS}} = {G \over Z(B_{\mu \nu})} \int {\cal D} B_{\mu \nu}
\exp \left[ -S ( B_{\mu \nu}) + i \int d^D x\ B_{\mu \nu} T_{\mu \nu} \right] \
,}
where $B_{\mu \nu}$ is an antisymmetric Kalb-Ramond field \ref\kal{V.I.
Ogievetsky and I.V. Polubarinov, Sov. J. Nucl. Phys. 4 (1967) 156;
M. Kalb and P. Ramond, Phys. Rev. D9 (1974) 2273; D.Z. Freedman and
P.K. Townsend, Nucl. Phys. B177 (1981) 282.}\ and 
\eqn\ddt{\eqalign{T_{\mu \nu}({\bf x}) &= {1\over 2} \int d^2\sigma\ X_{\mu
\nu}(\sigma)\ \delta^D({\bf x} - {\bf x}(\sigma)) \ , \cr
X_{\mu \nu} &= \epsilon^{ab} {\partial x_\mu \over \partial \sigma^a} 
{\partial x_\nu \over \partial \sigma^b} \ ,\cr}}
with ${\bf x}(\sigma)$\ parametrizing the world-sheet.
$G$ is a group factor which takes the value $G = 1$\ for compact Abelian gauge
theories and reduces at long distances to $G = N^{- \chi}$, with $\chi$\ the Euler characteristic, for
SU(N) gauge theories. The action for the Kalb-Ramond tensor is given by
\eqn\kra{S ( B_{\mu \nu}) = \int d^Dx \ f(H_{\mu \nu
\alpha}) + {1 \over 4 e^2} B_{\mu \nu} B_{\mu \nu} \ ,}
where $f$  takes the form
\eqn\fmv{\eqalign{f(H_\mu) &= {2\over 3}\ \Lambda_0 H_\mu\ {\rm sinh^{-1}}\left({2 \over
3 z \Lambda_0^3}\ H_\mu \right) - z \Lambda_0^4 \sqrt{1 + {4
\over 9 z^2 \Lambda_0^6} H_\mu^2} \ , \cr
H_\mu &\equiv \epsilon_{\mu \nu \alpha \beta} H_{\nu \alpha \beta}\ ,\cr
H_{\mu \nu \alpha} &= \partial_\mu B_{\nu \alpha} + \partial_\nu B_{\alpha \mu}
+ \partial_\alpha B_{\mu \nu} \ , \cr}}
in 4D.
Here $e^2$\ is the dimensionless coupling  of the original gauge theory,
$\Lambda_0$\ is the cutoff, which is needed in 4D, and $z \propto \exp (- {\rm
const}/e^2)$.

The action of the confining string was motivated by the facts that it can be
explicitly derived for compact U(1) theories, where it arises due to
the condensation of topological defects \amp, and that, up to as yet uncontroled 
contact terms,
it satisfies the string loop equations \amp.
 At long distances (and weak fields) the Kalb-Ramond action can be 
 approximated by 
 \eqn\akr{\eqalign{S ( B_{\mu \nu}) &= \int d^Dx\ {1 \over 12 \Lambda^2}\ H_{\mu \nu
\alpha} H_{\mu \nu \alpha} + {1 \over 4 e^2} B_{\mu \nu} B_{\mu \nu} \ ,\cr
\Lambda &= {\Lambda_0 \over 4} \sqrt{z}\ .\cr}}
In this form, the confining string action was independently proposed in
\ref\top{F. Quevedo and C.A. Trugenberger, CERN-TH/96-109, hep-th/9604196.},
where it was viewed as a special case of the generic Julia-Toulouse mechanism
for the confinement of $(h-1)$-branes by the condensation of $(D-h-3)$-branes in a
compact antisymmetric tensor field theory of rank $h$.
Eq.\akr\ corresponds to $h$ = 1 and $\Lambda$ is the new scale generated by the
condensation of $(D-4)$-branes, as explained in \top.
For example, in D=3, $\Lambda^3$\ is proportional to the average density of
instantons in Euclidean space.
The mass of the Kalb-Ramond field in \akr\ is then given by $m = \Lambda/e$.

The purpose of the present note is threefold.
First we woud like to present a full derivation of the confining string action
by performing an exact duality transformation; this was done only
at the semiclassical level in [9].
Secondly, we would like to investigate the issue of the extrinsic curvature term in
the low-energy limit of the confining string.
Finally, using the  formalism developed in \top, we would like
to study how the confining string action is modified by the presence of a
$\theta$-term,
\eqn\act{\eqalign{&S = \int d^4x\ {1 \over 4 e^2} F_{\mu \nu} F_{\mu \nu} + i
{\theta \over 64 \pi^2} F_{\mu \nu} \epsilon_{\mu \nu \alpha \beta}
 F_{\alpha \beta} \ ,\cr
&F_{\mu \nu} = \partial_\mu A_\nu - \partial_\nu A_\mu\ ,\cr}}
in the action for the Coulomb phase of a 4D compact U(1) gauge theory.

We shall find that, indeed, the extrinsic curvature term enters the (Euclidean)
action with a {\it negative\ coefficient}, thereby confirming the observation
of Kleinert \hke.
Note, however, that in the confining string the extrinsic curvature term appears
only in a long distance (low-energy) derivative expansion.
The full action is an induced action and thus automatically non-local as
suggested in \bra.
Moreover, we shall find that only for $\theta \neq 0$ one can take the limit
$\Lambda_0 \to \infty$, in which case $e$ is driven to infinity, $e \to \infty$.
The resulting action is the Nambu-Goto action plus a topological term measuring
the self-intersection number of the world-sheet.
This is in accordance with the original observation of Polyakov \pol, \amp\
that a $\theta$-term might stabilize the rigid string.

\newsec{Duality Transformation}

Starting with Euclidean $4D$ compact QED, the condensation of monopoles 
for $e>e_{cr}$ induces an effective action for the dual gauge field $\varphi_\mu$. Following
Polyakov [2] and Orland \ref\orland{P. Orland, Nucl. Phys. B205[FS5] (1982) 107.} , 
the corresponding partition function  can be written as:
\eqn\prim{\eqalign{Z&=\int{\cal D}\varphi_\mu\ \exp\left\{-S_{\rm conf}\right\}, \cr
S_{\rm conf}&=\int d^4x\, \left\{4e^2 f_{\mu\nu} f_{\mu\nu}+
z\Lambda_0^4\left(1-\cos{\varphi_\mu\over\Lambda_0}\right)\right\},\cr}}
with $f_{\mu\nu}\equiv \partial_\mu\varphi_\nu-\partial_\nu\varphi_\mu$.
This is the 4D extension of the well known 3D case\foot{
Everything we will say in this section applies for different
numbers of dimensions but we write it explicitly in 4D for concreteness.}.
However, in 4D, this partition function has
to be understood in terms of a lattice
regularization, as emphasized in [9]:
\eqn\latt{\eqalign{Z_l &= \int {\cal D}\varphi _{\mu }\ {\rm exp}
 \left( -S_l\right)
\ ,\cr
S_l &= \sum_{\bf x} \left( 4e^2l^4 f_{\mu \nu }f_{\mu \nu } + z\left( 1-{\rm cos}
(l\varphi_{\mu }) \right) \right) \ ,\cr }}  
with $l\equiv 1/\Lambda _0$, ${\cal D}\varphi_{\mu } \equiv \prod _{\bf x,\mu } d
\varphi _{\mu } ({\bf x})$ and $f_{\mu \nu }\equiv d_{\mu }\varphi _{\nu }-
d_{\nu }\varphi _{\mu }$, with $d_{\mu }$ the (forward) lattice derivative. The
UV cut-off $\Lambda _0=1/l$ is needed in 4D since the coupling $e$ is large.

In order to get the action $S(B_{\mu\nu})$, we should perform a duality transformation 
starting with the path integral
\eqn\tres{Z=\int {\cal D} B_{\mu \nu}{\cal D}\varphi_\mu\
\exp \left( -S ( B_{\mu \nu}, \varphi_\mu) \right) \ ,}
with the first-order Euclidean action
\eqn\seg{S(B_{\mu\nu},\varphi_\mu)=\int d^4x\, {1\over 4e^2} B_{\mu\nu}B_{\mu\nu}
+i\varepsilon_{\mu\nu\alpha\beta}B_{\mu\nu}f_{\alpha\beta}
+z\Lambda_0^4\left(1-\cos{\varphi_\mu\over\Lambda_0}\right).}
Integration over the field $B_{\mu\nu}$, being a Gaussian, can be easily performed and 
brings back $S_{\rm conf}$
as it was explicitly shown by Polyakov.
To find the dual action, which is the one describing the confining string, we 
would have to integrate out the field $\varphi_\mu$. 
Notice that the non-linear dependence on this field makes the 
integration seem essentially untractable.
Polyakov proceeded by eliminating $\varphi_\mu$ by its field equation in Minkowski space
and
substituting back into the action. In our 4D case this would give 
\eqn\mink{f_{\rm Mink}\left(H_{\mu }\right) = {2\over 3} \Lambda _0 H_{\mu }
{\rm arcsin}\left( {2\over 3z\Lambda _0^3} H_{\mu } \right) + z\Lambda _0^4
\sqrt{1-{4\over 9z^2\Lambda _0^6}H_{\mu }^2} \ ,}
which is the 4D extension of Polyakov's 3D result. The (Minkowski) partition function
implies a sum over the branches of the {\it multivalued} function $f_{\rm Mink}
\left( H_{\mu } \right)$. Polyakov \mpo \ showed how this sum over branches can be traded
for a summation over surfaces.

Here we shall instead derive the exact action in Euclidean space. We start by
rewriting the (lattice-regularized) path integral \tres \ as follows
\eqn\per{Z_l=\int {\cal D}B_{\mu \nu } \sum_{\{ n_{\mu } \} } \exp \left(2\pi i\sum_{\bf x}
l^3 B_{\mu \nu } t_{\mu \nu } \right) \int _{-\pi/l}^{+\pi/l} {\cal D}
\varphi_{\mu }\
\exp \left( -S\left( B_{\mu \nu }, \varphi _{\mu } \right) \right) \ ,}
with $t_{\mu \nu }\equiv \epsilon _{\mu \nu \alpha \beta} \left( d_{\alpha}n_{\beta}-
d_{\beta }n_{\alpha } \right)$.

In going from \tres \ to \per \ we have restricted the $\varphi _{\mu }$ integrations
to the fundamental domain at the price of introducing an additional set of link
variables $n_{\mu }$ which take into account all other periods of the $\cos$ function.

At this point we can perform the integrations over $\varphi _{\mu }$. To this end
we use the following result \ref\grad{I.S. Gradstheyn and I.M. Ryzhik, ``Table of Integrals, Series, and
Products'', Academic Press, Boston (1980).}:
\eqn\bess{{\rm e}^{a\cos x} = \sum_{k \in Z} I_k(a)\ {\rm e}^{ikx}\ ,}
with $I_k(a)$ a modified Bessel function. Using this we obtain the following contribution
to the partition function from the $\varphi_{\mu }$ integrations (up to an
irrelevant factor):
\eqn\phiint{\eqalign{Z_{\varphi } &= \prod_{\bf x, \mu } \int _{-\pi/l}^{+\pi/l}
d\varphi _{\mu }({\bf x})\ \exp \left( z\cos (l\varphi _{\mu } ) + i{2\over 3} l^4
\varphi _{\mu }H_{\mu } \right) \cr
&= \prod_{\bf x, \mu } \sum_{n_{\mu }(\bf x)} I_{n_{\mu}}(z) \int_{-\pi}^{+\pi}
d\varphi _{\mu }\ \exp i\left( {2\over 3}l^3 \varphi _{\mu }H_{\mu } - n_\mu \varphi _\mu
\right) \cr
&= \prod_{{\bf x}, \mu } \sum _{n_{\mu }({\bf x})} I_{n_{\mu }} (z)\ 
\delta _{n_{\mu }, {2\over 3} l^3 H_{\mu }} \cr
&=\prod _{{\bf x}, \mu }\ I_{{2\over 3}l^3H_{\mu }}(z) \ ,\cr }}
where we have absorbed $l$ in the definition of $\varphi_\mu$. Note that
Kronecker delta conditions imply
 the quantization condition $(2/3)l^3H_{\mu }(\bf x)= {\rm integer}$ for all
${\bf x}$ and $\mu $, which means that the unit of magnetic charge is quantized.
In the continuum limit we also get the total condition 
$\int _V d^3x {\bf H}\cdot {\bf n} =0$ where $V$ is any 3D hypervolume in 4D with
unit normal ${\bf n}$. This is an expression of the neutrality and isotropy of
the underlying monopole condensate.

For strong coupling, only small values of $\varphi_\mu $ contribute to the
partition function, as can be seen from \latt \ (note that an overall shift of
$\varphi _{\mu }$ by $2\pi n/l$ is irrelevant since the potential is periodic).
In this case it is a good approximation to restrict to values $|l t_{\mu \nu }|\le 1$.
In this `dilute gas approximation' the sum over configurations $\{ n_{\mu } \}$ reduces
to a sum over {\it closed surfaces} on the lattice. Restoring the continuum notation
with the appropriate factors of the UV cut-off $\Lambda _0$ we obtain
\eqn\strings{\eqalign{Z &= \int _{{\rm closed} \atop {\rm surfaces}} \int
{\cal D}B_{\mu \nu } \exp \left( -S\left( B_{\mu \nu }\right) +i \int _{\rm surface}
B_{\mu \nu }d\sigma _{\mu \nu } \right) \ ,\cr
S\left( B_{\mu \nu } \right) &= \int d^4x \ \left( -\Lambda_0^4 
\log I_{\left({2H_{\mu }
\over 3\Lambda_0^3}\right)}(z) + {1\over 4e^2} B_{\mu \nu }B_{\mu \nu } \right) \ ,\cr }}
i.e. the partition function of a string theory with action induced by the
Kalb-Ramond tensor field $B_{\mu \nu }$.

In the semiclassical approximation  the modified Bessel function $I_p(a)$ behaves
like \ref\bat{See for instance, Erd\'elyi {\it et al}, `Higher Trascendental Functions'
Vol. 2. McGraw Hill, New York (1953).}
\eqn\siet{\log I_p(a)\sim (p^2+a^2)^{1/2}-p \sinh^{-1}(p/a)\ .}
Using this in \strings \ we obtain the Euclidean version of Polyakov's result
presented in the previous section. Furthermore, for small $p/a$ the modified Bessel 
functions satisfy 
\eqn\och{{I_p(a)\over I_0(a)}\sim e^{-p^2/2a}\ .}
Therefore, the partition function reduces to the 
standard gaussian result \akr \ (after the overall factor of $I_0(a)$
is absorbed in the integral).

An analogous computation in Minkowski space leads to \mink . Note, however, that only
in Minkowski space the sum over surfaces is equivalent to a sum over branches of
a multivalued function, as stressed in \mpo . Our result 
indicates that the key feature of the dynamical
generation of strings is the periodic structure of the potential induced by the
condensation of the topological defects. Strings arise from the summation over the
periods of this potential in the duality transformation.

Note that this computation can be generalized. Indeed,  
the integration over $\varphi_\mu$ in equation 
\tres\ is a particular case of the following path integral
in $D$ dimensions:
\eqn\cuatr{ Z=\int {\cal D}A{\cal D}F\
\exp \left(i\int d^Dx \left(A \cdot (\partial F)\right)-\int d^Dx\left\{
g(A^2)+h(F^2)\right\}
\right)\ ,}
where $A$ and $F$ are  antisymmetric tensors of rank $r$ and $D-r-1$
respectively and $h$ and $g$ are arbitrary functions,
also $A \cdot (\partial F)\equiv \varepsilon_{\mu_1\cdots\mu_D}
A_{\mu_1\cdots\mu_r}\partial_{\mu_{r+1}}F_{\mu_{r+2}\cdots\mu_D}$.
Integration over $A$ gives precisely the  Fourier transform
of $e^{-g}$, therefore we can write:
\eqn\cinc{ Z=\int {\cal D}F\, \exp\left(-\int d^Dx\left\{\tilde{g}\left((\partial F)^2
\right)+h(F^2)\right\}\right)\ ,}
where $e^{-\tilde{g}}$ is the Fourier transform of
$e^{-g}$ \ref\kaw{H. Kawai, Progr. Theor. Phys., 65 (1981) 351}.
For instance, since the Fourier transform of $e^{-x^2}$ is $e^{-p^2/4  }$, 
we can easily recover the well known result that a massive tensor
of rank $r$ is dual to a massive tensor of rank $D-r-1$ in $D$ dimensions
($g(A^2)=A^2$ and $h(F^2)=F^2$).
We can then expect that for a theory of a massless antisymmetric tensor of rank
$r$, the condensation of topological defects of dimension $D-r-2$ gives rise to
a phase defined by `confining $r$-branes'.

\newsec{Low-Energy Effective Action and $\theta$-Term}
In a {\it compact} U(1) theory the $\theta$-term produces non-trivial effects,
notably it assigns an electric charge $q = e \theta/2 \pi$ to elementary
magnetic monopoles \ref\wit{E. Witten, Phys. Lett. B86 (1979) 283.}.
The confining phase arises thus due to the condensation of dyons.
Since we are mostly interested in the low-energy limit of the confining string
we shall use the quadratic expansion \akr\ of the Kalb-Ramond action.
Following \top\ the change induced by the $\theta$-term is given by
\eqn\mkr{S ( B_{\mu \nu}) = \int d^4x\ {1 \over 12 \Lambda^2} 
H_{\mu \nu \alpha} H_{\mu \nu \alpha} + {1 \over 4 e^2} B_{\mu \nu} B_{\mu
\nu}\
+ i {\theta \over 64 \pi^2} B_{\mu \nu} \epsilon_{\mu \nu \alpha \beta}
 B_{\alpha \beta} \ .}
Correspondingly, the mass of the Kalb-Ramond field is modified from $m = \Lambda/e$
to
\eqn\mas{\eqalign{m_\theta &= {e \Lambda \over 4 \pi} \sqrt{\left( {4 \pi \over
e^2}\right)^2 + t^2} \ ,\cr
t &\equiv {\theta \over 2 \pi}\ .\cr}}
This mass is determined by the same modular parameter $\tau = \left( \theta/2\pi
\right) + i\left( 4\pi/e^2 \right)$ which enters the mass formula for the dyons
in the BPS limit \ref\oli{For a recent review see: D. Olive, hep-th/9508089.}.
Note that, in the limit $\Lambda \to 0$ in which the density of monopoles
(dyons) vanishes, the action \mkr\ simply reduces to the action \act\ of pure
QED, since only configurations for which $H_{\mu \nu \alpha} = 0$ contribute to
the partition function in this case.
Correspondingly, the confining string action $S_{\rm CS}$ in \csa\ reduces to
the Coulomb interaction of the end-points of open strings:
\eqn\pop{\eqalign{\lim_{\Lambda \to 0} S_{\rm CS} &= - \ln \langle W(C)
\rangle_{\rm Coul} = {e^2 \over 2} \int d^4x\ j_\mu {1 \over - \nabla^2} j_\mu
\ ,\cr
j_\mu ({\bf x})  &\equiv 2 \partial_\nu T_{\mu \nu} ({\bf x})
 = \int_{\rm C} d\tau\ {d x_\mu \over d \tau}\ \delta^4 \left({\bf x} - 
 {\bf x}(\tau) \right) \ , \cr}}  
where C denotes the closed world-line bounding the original open world-sheet.

Performing the Gaussian integration over the field $B_{\mu\nu}$ we obtain the
confining string action in the form
\eqn\sdm{S_{\rm CS} = \int d^4x\  T_{\mu \nu} {\Lambda^2 \over m_\theta^2
-\nabla^2} T_{\mu \nu} + 2e^2 \partial_\nu T_{\mu \nu} {1 \over m_\theta^2
-\nabla^2} \partial_\alpha T_{\mu \alpha} + 
 i {e^2 \Lambda^2 \theta \over 16 \pi^2} T_{\mu \nu} {\epsilon_{\mu \nu
\alpha \beta} \over m_\theta^2
-\nabla^2} T_{\alpha \beta}\ .}
To further analyze the low-energy limit of \sdm\  we
first introduce the representation \ddt\ in \sdm, and then perform a derivative
expansion of the resulting action.
Notice that this is equivalent to sending the cut-off or, equivalently,
$m_\theta$, to $\infty$; in other words, the derivative expansion is equivalent
to an expansion in powers of $1/ m_\theta$.

We start by noting that the 4D Yukawa Green function in \sdm\ is given by
\eqn\fdg{G({\bf x}) \equiv {1 \over m_\theta^2 - \nabla^2} \delta^4({\bf x}) =
{m_\theta^2 \over 4 \pi^2} {1 \over m_\theta r} K_1 (m_\theta r)\ ,}
with $r \equiv |{\bf x}|$ and $K_1$ a modified Bessel function  \grad .
In computing \sdm\ using the representation \ddt\ we encounter the expression
$G\left( \sqrt{g_{ab} ({\bf \sigma}) \epsilon^a \epsilon^b} \right)$ with
\eqn\dva{g_{ab} = {\partial x_\mu \over \partial \sigma^a} 
{\partial x_\mu \over \partial \sigma^b}\ ,}
the {\it induced\ metric} on the world-sheet.
The derivative expansion of this Green function on the world-sheet is obtained
by expanding $\int d^2\epsilon\ G\left( \sqrt{g_{ab} \epsilon^a \epsilon^b}
 \right) f({\bf \epsilon })$ in powers of $1/m_\theta$ for any test
function $f({\bf \epsilon })$.
Naturally, the coefficients of this expansion may diverge and must then
 be regularized by
the ultraviolet cut-off $\Lambda_0$.
We obtain
\eqn\edg{\eqalign{G\left( \sqrt{g_{ab} \epsilon^a \epsilon^b} \right) &= {1
\over 2 \pi} g^{-1/2} K_0\left({m_\theta \over \Lambda_0}\right) \delta^2 ({\bf
\epsilon}) +  {1 \over 4 \pi m_\theta^2} g^{-1/2} g^{ab} \partial_a \partial_b
 \delta^2 ({\bf \epsilon}) + ... \ ,\cr
 g &= {\rm det} g_{ab} = {1 \over 2} X_{\mu \nu} X_{\mu \nu}\ .\cr}} 
Inserting this expression in \sdm\ we obtain the desired derivative expansion
of the confining string action $S_{\rm CS}$ up to terms of $O(m_\theta^0)$:
\eqn\iaf{\eqalign{S_{\rm CS} &= {\Lambda^2 \over 4 \pi} K_0\left({m_\theta
\over \Lambda_0} \right) \int d^2\sigma\ \sqrt{ g} - {\Lambda^2 \over 16 \pi
m_\theta^2} \int d^2\sigma\ \sqrt{ g} g^{ab} \partial_a t_{\mu \nu} \partial_b
t_{\mu \nu} + {\Lambda^2 \over 16 \pi
m_\theta^2} \int d^2\sigma\ \sqrt{ g} R  \cr &- i {\pi t \over \left( {4 \pi
\over e^2} \right)^2 + t^2}\ \nu + {e^2 m_\theta \over 8 \pi^2} f\left({m_\theta
\over \Lambda_0} \right) \int_{\rm boundary} d \tau\ \sqrt{{dx_\mu \over d\tau}
{dx_\mu \over d\tau}} + ... \ ,\cr}}
where 
\eqn\ndv{t_{\mu \nu} \equiv g^{-1/2} X_{\mu \nu}\ ,} 
and $R$ is the scalar curvature of the world-sheet.
The function $f\left({m_\theta \over \Lambda} \right)$ is defined as: 
$$f (x) = \int_x^\infty {dz \over z} K_1 (z)\ .$$
The first term in the expansion \iaf\ is the Nambu-Goto term.
The second and third term represent the {\it extrinsic} and intrinsic {\it
curvature} terms, respectively.
Note that the extrinsic curvature terms enters the Euclidean action with a {\it
negative\ coefficient}, as anticipated.
The fourth term is a topological term since
\eqn\qtt{\nu \equiv {1\over 4 \pi} \int d^2\sigma\ \sqrt{ g}\ 
\epsilon^{\mu \nu \alpha \beta}  
g^{ab} \partial_a t_{\mu \nu} \partial_b
t_{\alpha \beta}}
is the analytic expression for the (signed) {\it self-intersection\ number} of
the world-sheet.
Finally, the fifth term represents a boundary correction of $O(m_\theta)$.
All higher order terms are suppressed by negative powers of $m_\theta$ and are
thus infrared irrelevant.

\noindent{\it Case with $\theta=0$}

In the following we would like to study the renormalized action when the
cut-off $\Lambda_0$ is removed. Let us begin with the case $\theta = 0$.
In this case, \iaf\ becomes:
\eqn\nia{\eqalign{S_{\rm CS} &= {\Lambda^2 \over 4 \pi} K_0\left( {\sqrt{z}
\over 4 e} \right) \int d^2\sigma\ \sqrt{ g} - {e^2 \over 16 \pi} \int d^2\sigma\ \sqrt{ g} g^{ab} \partial_a t_{\mu \nu} \partial_b
t_{\mu \nu} + {e^2 \over 16 \pi} \int d^2\sigma\ \sqrt{ g} R  \cr &+ 
{e \Lambda \over 8 \pi^2} f\left( {\sqrt{z}
\over 4 e} \right) \int d \tau \sqrt{{dx_\mu \over d\tau}
{dx_\mu \over d\tau}} \ .\cr}}
This result is however obtained by a derivative expansion and is thus valid only on scales
much bigger than $1/m$, with $m = \Lambda / e$. Moreover, because we are in the
phase in which the monopoles (dyons) are condensed, our result is valid only
in the strong coupling limit $e > e_{\rm cr}$.
In the limit $\Lambda_0 \to \infty$ we can use it only if $e \to $ const or $e
\to \infty$ such that $m \to$ const or $m \to \infty$. However, in all these
cases the string tension diverges and the strings are suppressed since $
\lim_{e \to \infty} K_0 (\sqrt{z}/4e) = \lim_{e \to \infty} K_0 ({\rm
const}/e) = \infty$.
If we take the limit $e \to \infty$ for fixed cut-off, or if $\Lambda_0 = o(e)$,
 the mass $m \to 0$.
In this case we have to resort to the original expression \sdm.
In the case of fixed cut-off, and 
on distance scales much larger than $1/\Lambda$ this reduces to the action
\eqn\afc{S_{\rm CS} = \int d^4x\ \Lambda^2 T_{\mu \nu} {1 \over - \nabla^2}
 T_{\mu \nu}}
for {\it closed\ strings} (boundary terms are suppressed).
If we compute this expression for a sphere of radius $R \gg 1/\Lambda$ we obtain
$$S_{\rm CS}^R \propto \Lambda^2 R^2 \ln R\Lambda\ ,$$
which means that strings are logarithmically confined on scales larger than 
$1/\Lambda$.
In the case in which the cut-off $\Lambda_0 \to \infty$, also \afc\ diverges
and again strings are completely suppressed.
For $\theta = 0$ the confining string in 4D makes  thus sense only as a cut-off
theory for finite $e$.

\noindent{\it Case with $\theta\neq 0$}

In the case in which $\theta \neq 0$, instead, we have  $m_\theta = {e \Lambda \over 4 \pi} \sqrt{\left( {4 \pi \over
e^2}\right)^2 + t^2}$, and 
$$m_\theta \simeq {e \Lambda t \over 4 \pi}\qquad {\rm for}\ e \gg 1\ .$$
Since $m_\theta \to \infty$ for $\Lambda_0 \to \infty$, we can use the derivative
expansion which takes the form
\eqn\itf{\eqalign{S_{\rm CS} &= {\Lambda^2 \over 4 \pi} K_0\left({e t
\over 4 \pi} \right) \int d^2\sigma\ \sqrt{ g} - {1\over 16 \pi} \left( {4 \pi
\over et} \right)^2 \int d^2\sigma\ \sqrt{ g} g^{ab} \partial_a t_{\mu \nu} \partial_b
t_{\mu \nu} +\cr &+ {1\over 16 \pi} \left( {4 \pi
\over et} \right)^2 \int d^2\sigma\ \sqrt{ g} R - i {\pi t \over 
\left( {4 \pi \over e^2} \right)^2 + t^2} \nu + {e^3 t \Lambda \over 32 \pi^3} 
f\left({e t
\over 4 \pi} \right) \int d \tau \sqrt{{dx_\mu \over d\tau}
{dx_\mu \over d\tau}} \ ,\cr}}
for $e \gg 1$.
We now use the asymptotic behaviours for large $x$:
\eqn\ase{ K_0 (x) \simeq {{\rm e}^{-x} \over \sqrt{x}} \ , \quad f(x) \simeq
{{\rm e}^{-x} \over x^{3/2}}\ ,}
and we take the simultaneous limit $\Lambda_0 \to \infty$ and $e \to \infty$ so
that
\eqn\fst{{\Lambda^2 \over 4 \pi} {{\rm e}^{-et \over 4 \pi} \over \sqrt{et
\over  4 \pi}} = T\ ,}
with $T$ the physical string tension.
In this limit only closed surfaces survive and the action becomes
\eqn\act{S_{\rm CS} = T \int d^2\sigma\ \sqrt{ g} - i {\pi \over t }\ \nu\ .}
The topological correction term to the Nambu-Goto action can be considered as a
``{\it spin\ term}'' for the string, analogous to the topological spin term of 2D
point particles (self-intersection number of the world-line) \ref\pok{A.M.
Polyakov, Les Houches 1988 Lectures, ``Fields, Strings and Critical Phenomena''
E. Br\'ezin and J. Zinn-Justin eds., North-Holland, Amsterdam (1990).}.
Note that $t = 1$ for $\theta = 2\pi$ i.e. for a dyon condensate of elementary
charge $e$. In this case the weight of self-intersections in the partition
function is $(-1)^\nu$, which is the mechanism originally advocated by Polyakov
for the stabilization of rigid strings \pol\amp. 
The periodicity in $\theta$ is presumably recovered for $Z_{\rm N}$ theories
which contain also electric excitations with no magnetic charges \ref\cer{J.
Cardy and E. Rabinovici, Nucl. Phys. B 205 [FS5] (1982) 1; J. Cardy, Nucl.
Phys. B205 [FS5] (1982) 17.}.

The expression \fst\ for the string tension can be rewritten as
\eqn\nfs{\ln {\Lambda^2 \over 4 \pi T} = {et \over 4\pi} + {1 \over 2} \ln {et
\over 4 \pi} \ .}
For $e \gg 1$ the logarithm can be neglected with respect to the linear
function and we can write
\eqn\efl{e = {4 \pi \over t} \ln {\Lambda_0^2 \over 4 \pi T}\ .}
As usual we can express the cut-off in terms of the physical string tension $T$
and a reference scale $\mu$ as $\Lambda_0 = \sqrt{ 4 \pi} T/ \mu$ so that
$\Lambda_0 \to \infty$ corresponds to the infrared limit $\mu \to 0$.
This gives us the running coupling constant
\eqn\rcc{e(\mu) = {4 \pi \over t} \ln {T \over \mu^2}\ ,}
In the infrared $\mu \to 0$ we have $e \to \infty$. At a finite scale  this
equation determines the perturbative corrections, like the extrinsic curvature,
to the free string with topological term.

\bigbreak\bigskip\bigskip
\noindent {\bf Acknowledgements} 

\noindent C.A.T. is supported by a Profil 2 fellowship from the Swiss National
Science Foundation.

\listrefs
\end